\documentclass[11pt,a4paper]{article}
\usepackage{jheppub}
\usepackage{amsmath}
\usepackage[most]{tcolorbox}
\usepackage{dsfont}
\usepackage{ulem}
\usepackage{natbib}
\usepackage{xcolor}
\usepackage[hang,flushmargin]{footmisc}
\usepackage{tikz-cd}
\usepackage{enumitem}
\usepackage{amsfonts,amssymb, amscd,amsmath,latexsym,amsbsy,bm}
\usepackage{stmaryrd}
\usepackage{todonotes}
\usepackage{stackrel}
\usepackage{amsthm}

\usepackage{graphicx}
\newcommand{\imi}{\mathsf{i}}

\title{A comment on the solutions of the generalized Faddeev-Volkov model}

\author{Mehmet Dede}
\affiliation{
	 {Department of Physics, Bogazici University, 34342 Bebek, Istanbul, Turkey}\\[-0.5cm]

}

\emailAdd{mehmet.dede@boun.edu.tr}

\abstract{We consider two recent generalizations of the Faddeev-Volkov model, which is exactly solvable Ising-type lattice spin model. The first generalization based on using of the non-compact quantum dilogarithm over Pontryagin self-dual LCA group $\mathbb{R}\times \mathbb{Z}/N\mathbb{Z}$, and  another one constructed in a recent study via the gauge/YBE correspondence. We show that weight functions of these models obtained by different techniques are the same up to a constant.}


\begin{document}
	\maketitle
	\flushbottom

\section{Introduction}
In recent years, there have been several progress in constructing new integrable systems in terms of Yang-Baxter equations and generalizing known solutions. Some studies employ the gauge/YBE correspondence to connect supersymmetric gauge theories and integrable lattice models of statistical mechanics, e.g. \cite{,Spiridonov:2010em,Yamazaki:2012cp,Gahramanov:2015cva,Gahramanov:2016ilb,Gahramanov:2017ysd,Kels:2017vbc,Yamazaki:2018xbx,Bozkurt:2020gyy,de-la-Cruz-Moreno:2020xop,Mullahasanoglu:2021xyf,Gahramanov:2022qge}, whereas others use special functions and algebraic techniques, see, e.g. \cite{Bazhanov:2010kz,Kels:2013ola,Kels:2015bda,Kashaev:2015nya,Andersen:2014aoa,Gahramanov:2017idz}. It turns out that, under the particular case, solutions derived from using two different constructions might take similar forms \cite{Eren:2019ibl}.

In the work \cite{Kashaev:2015nya}, Kashaev generalized the solution of the Faddeev-Volkov statistical model, utilizing the quantum dilogarithm $\phi(x,m)$ over LCA group $A$, and the Boltzmann weight function
\begin{equation}\label{weight}
     W(x,y)=\frac{\phi(x-y)}{\phi(x+y)}\langle x;y \rangle = \frac{\phi(x-y) \phi(-x-y)}{\phi(0)^2 \langle x \rangle \langle  y\rangle} \; ,
\end{equation}
where $x$ and $y$ take values in specified self-dual LCA group $A$, and $\langle \cdot \rangle : A \rightarrow \mathbb{T}$ is Gaussian exponential, whose existence is implied by the definition of self-duality of LCA group \cite{Kashaev:2015nya}. The weight function (\ref{weight}) solves the operator Yang-Baxter equation
\begin{equation}
    W(\textbf{p},x) W(\textbf{q},x+y)W(\textbf{p},y)=W(\textbf{q},y) W(\textbf{p},x+y)W(\textbf{q},x)\; ,
\end{equation}
where \textbf{p} and \textbf{q} are self-adjoint operators in Hilbert space and they act as differentiation and multiplication operators respectively. Therefore, one can construct the corresponding integrable lattice spin model of statistical mechanics \cite{Baxter:1982zz}. 

We set our framework around a particular case $A=\mathbb{R}\times \mathbb{Z}/N\mathbb{Z}$. Over LCA group $\mathbb{R}\times \mathbb{Z}/N\mathbb{Z}$ the Gaussian exponential takes the form $\langle (x,m) \rangle=e^{-\pi \imi x^2 }e^{-\pi \imi m(m+N)/N}$ and explicit version of quantum dilogirthm $\phi(x,m)$  can be written as \cite{Andersen:2014aoa,Kashaev:2015nya}:
\begin{equation}\label{quantumdilog}
\phi(x,m)=\prod_{j=0}^{N -1}\Phi_b\left(\frac{x}{\sqrt{N }}+\imi(1-N^{-1})\left(\frac{b+b^{-1}}{2}\right)-\imi b^{-1}\frac{j}{N }-\imi b\left\{\frac{j+m}{N }\right\}\right) \;,
\end{equation}
where $\{x\}$ represents the fractional part of $x$. 

On the other hand, recently, in \cite{Bozkurt:2020gyy} there was obtained a new solution to the star-triangle equation (a special form of the Yang-Baxter equation), which is a generalization of the Faddeev-Volkov model \cite{Bazhanov:2007mh,Bazhanov:2007vg} using gauge/YBE correspondence. Namely in order to construct a new solution to the star-triangle equation, authors of \cite{Bozkurt:2020gyy} used three-dimensional supersymmetric duality for $\mathcal N=2$ theories on lens space $S^3/\mathbb{Z}_r$.  

In this work, we show that two generalizations of the Faddeev-Volkov model constructed in \cite{Kashaev:2015nya} and \cite{Bozkurt:2020gyy} are identical in regard to their integrability properties.

\section{Hyperbolic gamma function and its properties}

%First, we will define some of the special functions that will be of use and provide the relations between them. 
To define these special functions, we need to introduce the q-Pochhammer symbol
\begin{align}
    (z,q)_\infty=\prod_{i=1}^\infty (1-z q^i), &&  \vert q\vert< 1 \;, && z \in \mathbb{C}\;.
\end{align}
Defining the q-Pochhammer symbol, we can write an explicit version of the non-compact quantum dilogartihm, which is the function Kashaev uses in his work \cite{Kashaev:2015nya} while generalizing the Faddeev-Volkov's solution \cite{Bazhanov:2007mh, Bazhanov:2007vg},
\begin{align}
    \Phi_b(z)= \frac{(-q_b e^{2\pi \imi b x};q^2_b)_\infty}{(-\Bar{q}_b e^{2 \pi \imi b^{-1} x};\Bar{q}_b^2)_\infty},  && q_b=e^{\pi \imi b^2 }, && \Bar{q}_b=e^{-\pi \imi b^{-2}}, && Im(b^2)>0 \;.
\end{align}

There is a similar function called the hyperbolic gamma function\footnote{This function appears in several areas of mathematical physics, this form of notation is often found in works of supersymmetry community, see, e.g. \cite{Dolan:2011rp,Spiridonov:2014cxa,Gahramanov:gka, Gahramanov:2013xsa, Gahramanov:2015tta,Bozkurt:2018xno}.} $\gamma^{(2)}(z,b,b^{-1})$. Using the q-Pochhammer symbol
\begin{align}
    	\gamma^{(2)}(z;b,b^{-1})=e^{-\frac{\pi i}{2}B_{2,2}(z;b,b^{-1})}\frac{(e^{2\pi i b}q^2_b; q^2_b)}{(e^{2\pi i\frac{z}{b}};\Bar{q}_b^2)}  \;,
\end{align}
where $B_{2,2}$ is the second diagonal Bernoulli polynomial,
\begin{equation}
    B_{2,2}(z;b,b^{-1})=z^2-z(b+b^{-1})+\frac{b^2+3+ b^{-2}}{6 }\;.
\end{equation}
The relation between hyperbolic gamma function $\gamma^{(2)}(z,b,b^{-1})$ and non-compact quantum dilogarithm $\Phi_b(z)$ can be stated as
\begin{equation}\label{eqeq}
    \Phi_b(z)=e^{-\pi \imi B_{2,2}(\frac{1}{2}(b+b^{-1})+iz,b,b^{-1})}\gamma^{(2)}(\frac{1}{2}(b+b^{-1})+iz;b,b^{-1})\;.
\end{equation}

One can also express the hyperbolic gamma function in integral form
\begin{equation}
   \gamma^{(2)}(z;b,b^{-1})=\exp{\left(-\int_{0}^{\infty}\frac{dx}{x}\left[\frac{\sinh{x(2z-b-b^{-1})}}{2\sinh{(xb)}\sinh{(xb^{-1})}}-\frac{2z-b-b^{-1}}{2x}\right]\right)} \;,
\end{equation}
where $Re(b),Re(b^{-1})>0 $ and $Re(b+b^{-1})> Re(z)>0$.

The hyperbolic gamma function has several practical properties, however, we will only use two of them:
\begin{align}\label{inv}
\gamma^{(2)}(z;b,b^{-1})\gamma^{(2)}(b+b^{-1}-z;b,b^{-1})=1 \\
	\label{scaling}
\gamma^{(2)}(z;b,b^{-1})=\gamma^{(2)}(u z;u b,u b^{-1})\;.
\end{align}

\section{Star-triangle relation}

In this section, we show that Kashaev's weight function differs from the Boltzmann weights constructed in \cite{Bozkurt:2020gyy} only by a scaling factor. To this end, we express $\Phi_b(z)$ in terms of the hyperbolic gamma function (\ref{eqeq}) and exploit the mentioned properties of the gamma function.

First, we replace the non-compact quantum dilogarithm $\Phi_b(z)$ with the hyperbolic gamma functions in the expression (\ref{quantumdilog})
\begin{align} \nonumber
    \phi(x,m)=\exp{\left(-\pi \imi \sum\limits_{j=0}^{N-1} B_{2,2}\left(\frac{\imi x}{\sqrt{N}}+\frac{(b+b^{-1})}{2}N^{-1}+\frac{b^{-1}j}{N}+b \left\{\frac{j+m}{N }\right\};b,b^{-1}\right)\right)}\\
    \prod_{j=0}^{N-1}\gamma^{(2)}\left(\frac{\imi x}{\sqrt{N}}+\frac{(b+b^{-1})}{2}N^{-1}+\frac{b^{-1}j}{N}+b \left\{\frac{j+m}{N }\right\};b,b^{-1}\right)\;.
\end{align}

In order to manipulate and use the properties of $\gamma^{(2)}(z,b,b^{-1})$, one may want to eliminate the fractional part in the argument. Thus, considering the fractional part, it can easily be seen that the product can be rewritten as two separated parts (for the simplicity let us denote $ \frac{\imi x}{\sqrt{N}}+\frac{(b+b^{-1})}{2}N^{-1}$  as $ \frac{y}{N}$)
\begin{align} \nonumber 
    \prod_{j=0}^{N-1}\gamma^{(2)}\left(\frac{y}{N}+\frac{b^{-1}j}{N}+b \left\{\frac{j+m}{N }\right\};b,b^{-1}\right)=\prod_{j=0}^{m-1}\gamma^{(2)}
\left({\frac{y}{N} }+b\left(1-{\frac{m}{N}}\right)+(b+b^{-1}){\frac{j}{N}};b,b^{-1}\right)\\ \label{productN}
\times\prod_{j=0}^{N-m-1}\gamma^{(2)}
\left({\frac{y}{N}}+{\frac{m}{N}}b+(b+b^{-1}){\frac{j}{N}};b,b^{-1}\right)\ \;.
\end{align}

Considering the product of two standard hyperbolic gamma functions in different bases, and using the limiting behavior mentioned in \cite{2018}, it can be shown that the expression on the right-hand side (\ref{productN}) is indeed equal to the product of two hyperbolic gamma functions. For the derivation of this relation and further relations, see \cite{Gahramanov:2016ilb,2018,Imamura:2012rq,Nieri:2015yia} 
\begin{align}\nonumber
    \prod_{j=0}^{N-1}\gamma^{(2)}\left(\frac{y}{N}+\frac{b^{-1}j}{N}+b \left\{\frac{j+m}{N }\right\};b,b^{-1}\right)=\gamma^{(2)}\left({y+m \frac{b}{N}};b,{\frac{b+b^{-1}}{N}}\right)\\
\gamma^{(2)}\left({\frac{y+(N-m)b^{-1}}{N}};b^{-1},{\frac{b+b^{-1}}{N}}\right)\;.
\end{align}

Therefore, the quantum dilogarithm $\phi(x,m)$ can be written as
\begin{align}\nonumber 
    \phi(x,m)=\exp{\left(-\pi \imi \sum\limits_{j=0}^{N-1} B_{2,2}\left(\frac{\imi x}{\sqrt{N}}+\frac{(b+b^{-1})}{2}N^{-1}+\frac{b^{-1}j}{N}+b \left\{\frac{j+m}{N }\right\};b,b^{-1}\right)\right)}
    \\\gamma^{(2)}\left(\frac{\imi x}{\sqrt{N}}+\frac{(b+b^{-1})}{2}N^{-1}+{\frac{m b}{N}};b,{\frac{b+b^{-1}}{N}}\right)\nonumber
    \\\label{summ}
 \gamma^{(2)}\left(\frac{\imi x}{\sqrt{N}}+\frac{(b+b^{-1})}{2}N^{-1}+{\frac{(N-m)b^{-1}}{N}};b^{-1},{\frac{b+b^{-1}}{N}}\right) \;.
\end{align}
Calculating the summation is nearly an identical task with calculating the product. One should separate the summation into two parts from $j=0$ to $m-1$ and from $0$ to $N-m-1$. Rest of the calculation follows from the definition of second diagonal Bernoulli polynomial $B_{2,2}(z, b,b^{-1})$ and elementary summation identities. After computing the summation, we insert the spin values which take values in LCA group. In Kashaev's setting, there are two different inputs of the weight function (\ref{weight}). The variable $x$ stands for differences of discrete spin ($m_1$ and $m_2$) and continuous spin ($x_1$ and $x_2$), hence $x=(x_1-x_2,m_1-m_2)$. The variable $y$ denotes the rapidity lines $(-\alpha,0)$.

Inserting these variables, we use scaling property of the hyperbolic gamma function (\ref{scaling}) where $u= -\imi N $. Finally, after the following change of variables
\begin{gather}
    x_1-x_2-\alpha \rightarrow -\imi \frac{x_1-x_2-\alpha}{\sqrt{N}}+\imi \frac{b+b^{-1}}{\sqrt{N}} \;, \\
    -x_1+x_2-\alpha \rightarrow -\imi \frac{x_1+x_2-\alpha}{\sqrt{N}}+\imi \frac{b+b^{-1}}{\sqrt{N}}\;,
\end{gather}
we observe that the weight function in Kashaev's settings is indeed a scaled version of the Boltzmann weight constructed in \cite{Bozkurt:2020gyy}
\begin{align}\nonumber
    W(x,y)=W_\alpha (x_1,x_2,m_1,m_2)=C e^{-\pi \imi (m_1 +m_2)}\gamma^{(2)}\left(-\imi (-\alpha +x_1-x_2)-\imi b(m_1-m_2)  ; -\imi b N , -\imi(b+b^{-1})\right)\times\\\nonumber
\gamma^{(2)}\left(-\imi (-\alpha+x_1-x_2)-\imi b^{-1} (N-(m_1-m_2));-\imi N b^{-1},-\imi(b+b^{-1}) \right)\times\\ \nonumber \gamma^{(2)}\left(-\imi (-\alpha-x_1+x_2)-\imi b(-m_1+m_2)  ; -\imi b N , -\imi (b+b^{-1})\right)\times\\\nonumber
\gamma^{(2)}\left(-\imi (-\alpha-x_1+x_2)-\imi b^{-1} (N-(-m_1+m_2)) ;-\imi N b^{-1},-\imi (b+b^{-1}) \right) \;,
\end{align}
with $C=\exp{\left[\frac{i \pi  b^2}{24 N}+\frac{i \pi }{24 b^2 N}-\frac{i \pi  N}{12}+\frac{i \pi }{12 N}\right] }$. Even though two Boltzmann weights differ from each other by a constant factor, they possess identical integrability properties as they satisfy the same star triangle equation 
\begin{align*}
    \frac{1}{\imi N}\sum\limits^{\lfloor N/2 \rfloor}_{u=-\lfloor N/2 \rfloor} e^{-2 \pi \imi u } \int\limits^\infty_{-\infty} dz W_{\alpha_1}(x_1,z, m_1,u) W_{\alpha_2}(x_1,z, m_1,u) W_{\alpha_3}(x_1,z, m_1,u)\\
    =W_{\alpha_1+\alpha_2}(x_1,x_2, m_1,m_2)W_{\alpha_1+\alpha_3}(x_1,x_2, m_1,m_2)W_{\alpha_2+\alpha_3}(x_1,x_2 ,m_1,m_2) \;.
\end{align*}

There are many solutions to the star-triangle relation obtained via the gauge/YBE correspondence, it would be interesting to study them from the algebraic point of view as in \cite{Kashaev:2015nya}.   

\bibliographystyle{utphys}
\bibliography{pentagon}

\end{document}